\documentstyle[12pt,epsfig]{article}

\newcommand{\gsim}{\stackrel{>}{\sim}}
\newcommand{\lsim}{\stackrel{<}{\sim}}

\newcommand\MeV{\,\mbox{MeV}}
\newcommand\GeV{\,\mbox{GeV}}

\begin{document}
\thispagestyle{empty}

\mbox{}
\begin{flushright}
DESY  94--044\\
March 1994\\
\end{flushright}
\vspace*{\fill}
\begin{center}
{\LARGE\bf ${\cal O}(\alpha^2 L^2)$ radiative corrections to
deep inelastic}

\vspace{2mm}
{\LARGE\bf $ep$ scattering for different kinematical }

\vspace{2mm}
{\LARGE\bf variables}
\\

\vspace*{20mm}
\large
\begin{tabular}[t]{c}
Johannes Bl\"umlein
\\

\vspace{2em}
\\

{\it DESY--Institut
f\"ur Hochenergiephysik, Zeuthen,}
\\
{\it Platanenallee 6,
D--15735 Zeuthen,  Germany}
\\
\end{tabular}
\end{center}
\vspace*{\fill}
\begin{abstract}
\noindent
The QED radiative corrections are calculated in the leading log
approximation up to ${\cal O}(\alpha^2)$ for different definitions of
the kinematical variables using jet measurement, the 'mixed' variables,
the double angle method, and a measurement based on $\theta_e$ and
$y_{JB}$. Higher order contributions due to
exponentiation of soft radiation are included.
\end{abstract}
\vspace*{\fill}
\newpage
\noindent
\section{Introduction}
The measurement of deep inelastic $ep$ scattering at HERA will allow
to extend the kinematical range to $x \sim 10^{-4}$ and
$Q^2 \sim  {\cal O}(10^4) \GeV^2$~\cite{KIN}. Because of differences
in the detector response and resolution \cite{RES} different ways to
measure the kinematical variables $x, y$ and $Q^2$ have to be used to
cover the full kinematical range. Furthermore, owing to the different
properties of detectors
 the experiments
H1 and ZEUS \cite{EXP} apply 
different methods in measuring the kinematical variables.

The QED radiative corrections in ${\cal O}(\alpha)$
differ significantly for the various
choices of kinematical variables as shown in \cite{RC}. This is due
to
the full (or partial) integration of the radiated photon's phase
space for different kinematical situations according to the
experimental requirements. 
As shown in
 explicit comparisons between
complete and leading log calculations in ${\cal O}(\alpha)$ 
the results agree to the per cent level
for the case of lepton measurement \cite{RCE1,RCE2},
jet measurement \cite{RCJ1,RCJ2}, and the case of
mixed variables
\cite{RCJ1,RCM1}~\footnote{
Also the case of 'hadronic' variables was investigated \cite{RCM1,RCH1}.
However, these
variables are not accessible in deep inelastic scattering
experiments.}. In part of the phase space the ${\cal O}(\alpha)$ 
corrections are sizeable. Therefore, it is necessary
 to calculate
the 2nd order corrections  to check whether the results obtained
are
numerically stable.

In this letter the ${\cal O}(\alpha^2)$ contributions are calculated
in the leading log approximation for the case of jet measurement,
mixed variables, the double angle method~\cite{DOUB}, and a measurement
based on $\theta_e$ and $y_{Jacquet-Blondel} \equiv y_{JB}$~\footnote{
For leptonic
variables these corrections have been calculated in \cite{RCE2X}.}.
The latter 
two methods have been applied by the ZEUS collaboration
recently in
measuring the deep inelastic neutral current cross section
\cite{SCHLE}. Since the ${\cal O}(\alpha L)$ radiative corrections
for these cases have not been given before the corresponding numerical
results will be included also.
Besides of the
second order leading log corrections
higher  order corrections due to soft exponentiation
are  given.

\section{${\cal O}(\alpha^2 L^2)$ corrections}
The relevant contributions arise from the
set of all possible collinear configurations in the diagrams up to
${\cal O}(\alpha^2)$. They are associated with the 'massless' fermion
lines ($m_e^2, m_q^2 \ll Q^2$). One may group the different processes
into:
\begin{itemize}
\begin{enumerate}
\item{bremsstrahlung (diagrams {\sf 1a,b}, figure 1)}
\item{$e^+e^-$ pair creation (diagram {\sf 1c}), and }
\item{$f\bar{f}$ pair production, $f = l^-,u,d,s,c,b$, (diagram
{\sf 1d}).}
\end{enumerate}
\end{itemize}
Among the possible configurations contributions due to initial state
and final state radiation both from the lepton and quark lines
emerge. Furthermore, the Compton scattering of a nearly real intermediary
photon, emitted from a quark or proton line and
scattered off the electron,
\cite{RCE1}
may be
considered as a part of the radiative corrections. 
Since the
experimental signature of the latter process consists 
of two nearly
balanced electromagnetic showers accompanied by 
low hadronic activity
near the beam pipe,
these events are generally not 
included into
 the deep inelastic sample.
On the other hand, they may be used to measure the proton
structure functions
at small $x$ and $Q^2$~\cite{COMP}.
Because of the finite detector resolutions
collinear final state radiation is  hardly resolved
and may be
treated integrally~\cite{KLN} in the experimental measurement.
Due to this also  terms with both collinear initial and final
state radiation do not emerge.
The initial state radiation from the quark line can be summed using
Altarelli--Parisi equations \cite{RUJU} with kernels modified
due to the 1st and 2nd order QED corrections, which yields a modification
of the parton distributions at the  level of 1\%~\footnote{Note, that
the current uncertainty in the parton distributions is larger than this
in most of the kinematical range \cite{MRSD}--\cite{GRV}.}.

Because of this, we will consider 
only the effect of
initial state radiation in the following.
For the scattering cross section up to
${\cal O}(\alpha^2)$ one thus obtains:
\begin{eqnarray}
\label{eqMA}
\frac{d^2 \sigma}{dx dy} &=&
\frac{d^2 \sigma^{(0)}}{dx dy} +
\frac{d^2 \sigma^{(1)}}{dx dy} +
\frac{d^2 \sigma^{(2)}}{dx dy} \nonumber\\
\frac{d^2 \sigma^{(1)}}{dx dy} &=&
 \frac{\alpha}{2 \pi}
\ln \left( \frac{Q^2}{m_e^2} \right) \int_0^1 dz
P_{ee}^{(1)}
(z) \left \{
\theta(z - z_0) {\cal J}(x,y,z)
\frac{d^2 \sigma^{(0)}}{dx dy} \left|_{x=\hat{x},
y=\hat{y},s=\hat{s}} \right. -  \frac{d^2 \sigma^{(0)}}{dx dy} \right \}
\nonumber\\
\frac{d^2 \sigma^{(2)}}{dx dy} &=&
 \frac{1}{2} \left[ \frac{\alpha}{2 \pi}
\ln \left( \frac{Q^2}{m_e^2} \right) \right]^2
\int_0^1 dz
 P_{ee}^{(2,1)}(z)
\left \{
\theta(z - z_0) {\cal J}(x,y,z)
\frac{d^2 \sigma^{(0)}}{dx dy} \left|_{x=\hat{x},
y=\hat{y},s=\hat{s}} \right. -  \frac{d^2 \sigma^{(0)}}{dx dy} \right \}
\nonumber
\end{eqnarray}
\begin{equation}
+
\left( \frac{\alpha}{2 \pi} \right )^2
\int_{z_0}^1 dz
 \left \{
\ln^2 \left( \frac{Q^2}{m_e^2} \right)
P_{ee}^{(2,2)}(z) +  \sum_{f=l,q} \ln^2 \left( \frac{Q^2}{m^2_f} \right )
P_{ee,f}^{(2,3)}(z) \right \}
{\cal J}(x,y,z)
\frac{d^2 \sigma^{(0)}}{dx dy} \left|_{x=\hat{x},
y=\hat{y},s=\hat{s}} \right.
\end{equation}
using standard  techniques  known from QCD~\footnote{These methods
have also been used for the calculation of QED corrections
to $e^+e^-$ annihilation processes successfully~\cite{EPEM}.}.
Here, $d^2\sigma^{(0)}/dx dy$ denotes the Born cross section
for deep inelastic neutral or charged current 
$ep$ reactions\footnote{Note, that
QED loop effects, as e.g. the polarization
of the intermediary boson, are not included in (\ref{eqMA}).
One may easily
account for them using  'dressed' Born cross sections
instead of $d \sigma^{(0)}/ dx dy$.}
 (cf. e.g.
(3,4) in \cite{RCJ1}), 
and  the Jacobian ${\cal J}$ is given by
\begin{equation}
{\cal J}(x,y,z) = \left | \begin{array}{cc}
\partial\hat{x}/ \partial x &
\partial\hat{y}/ \partial x   \\
\partial\hat{x}/ \partial y &
\partial\hat{y}/ \partial y
\end{array} \right |
\end{equation}
The shifted variables $\hat{x} \equiv \hat{Q}^2/\hat{y} \hat{s},~\hat{y},
 \hat{Q}^2$, $\hat{s}$,
and the lower bounds of $z$, $z_0$, are
listed in table~1 for the different choices of measurement.

Both for the double angle method and the measurement based on
$\theta_e$ and $y_{JB}$, the rescaled variables $\hat{x}$ and
$\hat{Q}^2$ vanish for $z \rightarrow z_0$ (cf. table~1).
If {\it no} futher cut is imposed  this means that for both methods
the radiative corrections are strongly influenced by the
non--perturbative range $x, Q^2 \rightarrow 0$ in
$d^2 \sigma/ dx dy$.
In studies by the ZEUS collaboration, where these two methods
have been applied, however, the cut
\begin{equation}
2 E_e = E'_e (1 - \cos\theta_e) 
+ E_J (1 -  \cos\theta_J) \geq {\cal A}
\end{equation}
on the reconstructed electron beam energy was imposed, choosing
${\cal A} = 35 \GeV$. Here,
$\theta_{e,J}$ denote
the scattering angle of the electron and the
jet angle, respectively, and  $E_e, E'_e, E_J$ are
the energies of the initial
electron, scattered electron, 
and the jet in the laboratory frame.
This cut yields 
$z_0 = {\cal A}/(2 E_e)$ 
in the case of the
double angle method and $z_0 = \max \{{\cal A}/(2 E_e),y \}$
for the measurement based on $\theta_e$ and $y_{JB}$.
For the double angle method the cut excludes the contributions
of the range $x, Q^2 \rightarrow 0$ globally, whereas they
are still present in the case of the $\theta_e-y_{JB}$
measurement for ${\cal A}/(2 E_e)~\lsim~y$.

The splitting functions in (\ref{eqMA}) are:
\begin{equation}
P_{ee}^{(1)}(z) = \frac{1 + z^2}{1 - z}
\end{equation}
\begin{eqnarray}
P_{ee}^{(2,1)}(z) &=& \frac{1}{2}
\left [P_{ee}^{(1)} \otimes P_{ee}^{(1)} \right ](z)
\nonumber\\
&=& \frac{1 + z^2}{1 - z} \left [ 2\ln(1-z) - \ln z + \frac{3}{2} \right ]
+ \frac{1}{2}(1 + z) \ln z - (1 -z)
\end{eqnarray}
\begin{eqnarray}
P_{ee}^{(2,2)}(z) &=& \frac{1}{2}
\left [
P_{e \gamma}^{(1)} \otimes P_{\gamma e}^{(1)}
\right ](z)  \nonumber\\
&\equiv&
(1 + z) \ln z +
\frac{1}{2}(1 - z) + \frac{2}{3} \frac{1}{z} (1 - z^3)
\end{eqnarray}
\begin{equation}
\label{eq-p23}
P_{ee,f}^{(2,3)}(z) =
N_c(f) e_f^2
\frac{1}{3} P_{ee}^{(1)}(z)
                        \theta \left ( 1 - z - \frac{2 m_f}{E_e}
                        \right )
\end{equation}
\noindent
Here, $P_{e \gamma}^{(1)}$ and $P_{\gamma e}^{(1)}$ are the
splitting functions of a photon into an electron, and vice versa.
$\otimes$ denotes the Mellin convolution.
$e_f$ is the fermion charge, and
$N_c(f) = 3$ for quarks, $N_c(f) = 1$ for leptons, respectively.
 (\ref{eq-p23})
is obtained from an expansion of $\alpha_{QED}(s)$ calculating the
polarization operator in the on--mass--shell scheme for leptons {\it and}
quarks, which can not be done
reliably in perturbative QCD in the case of the light
quarks. However, one may try to {\it parametrize}
$\alpha_{QED}(s, m_{f_i})$
in terms of {\it effective} quark masses  from a fit to
$R(s) = \sigma(e^+e^- \rightarrow {\rm hadrons})/
\sigma(e^+e^- \rightarrow \mu^+\mu^-)$ as done in~\cite{FJ}.
We will apply this description here and
use the parameters
$m_u =  62 \MeV$, $m_d =  83 \MeV$, $m_s = 215 \MeV$,
$m_c = 1.5 \GeV$, and $m_b = 4.5 \GeV$ as
obtained in \cite{FJ}
in the subsequent analysis.

The soft exponentiation beyond the second order can be calculated
using the solution of the non--singlet evolution equation for the
electron or positron distribution in the soft range 
($z \rightarrow 1)$ for
running $\alpha_{QED}$. It is
given by~\cite{GL}
\begin{equation}
\label{eqNS}
D_{NS}(z,Q^2) = \zeta (1 - z)^{\zeta - 1}
\frac{\exp\left [ \frac{1}{2} \zeta \left (
\frac{3}{2} - 2 \gamma_E \right ) \right ]} {\Gamma(1 + \zeta)}
\end{equation}
with
\begin{equation}
\label{eqzet}
\zeta = -3 \ln \left [ 1 - (\alpha/3\pi) \ln(Q^2/m_e^2) \right ]
\end{equation}
Expanding (\ref{eqNS}) up to ${\cal O}(\alpha^2)$, one obtains
terms in the limit $z \rightarrow 1$  which are
contained in (\ref{eqMA}) already.
Apart from bremsstrahlung terms also the $e^+e^-$ pair production
term  in (\ref{eq-p23}) is included
for $m_e \rightarrow 0$~\footnote{One might
generalize (\ref{eqzet}) using
also the heavier fermions $u, d, \mu, s,...$ in the running
coupling constant. However, apart from of the mass threshold in
(\ref{eq-p23}) the smaller logarithm associated with these terms
will yield an even smaller contribution.}.
To account  for the terms of ${\cal O}(\alpha^3)$ and higher  orders
contained in (\ref{eqNS}), we use
\begin{equation}
\label{eqP3}
P_{ee}^{>2,~soft}(z,Q^2) = D_{NS}(z,Q^2) -
\frac{\alpha}{2 \pi} \ln \left ( \frac{Q^2}{m_e^2} \right )
\frac{2}{1 - z} \left \{ 1 +
\frac{\alpha}{2\pi} \ln \left ( \frac{Q^2}{m_e^2} \right )
\left [ \frac{11}{6} + 2 \ln(1 - z) \right ] \right \}
\end{equation}
and\footnote{To obtain consistency between
eqs. (\ref{eqP3}) and (\ref{eqMA}) the limit
$m_e \rightarrow 0$  in
(\ref{eq-p23}) is taken.}
\begin{equation}
\frac{d^2 \sigma^{(>2,soft)}}{dx dy} =
 \int_0^1  dz
P_{ee}^{(>2)}(z,Q^2) \left \{
\theta(z - z_0) {\cal J}(x,y,z)
\frac{d^2 \sigma^{(0)}}{dx dy} \left|_{x=\hat{x},
y=\hat{y},s=\hat{s}} \right. -  \frac{d^2 \sigma^{(0)}}{dx dy} \right \}
\end{equation}

Beginning with the second order corrections in $\alpha$ terms
due to $e^- \rightarrow e^+$ conversion in the initial state
contribute. They result from diagram~{\sf 1c} in 
${\cal O}(\alpha^2)$ where now the outgoing fermion collinear
with the incoming electron is a positron.
The conversion rate equals
\begin{equation}
P(z,Q^2; e^- \rightarrow e^+) = \left(\frac{\alpha}{2 \pi} \right )^2
\ln^2 \left ( \frac{Q^2}{m_e^2} \right ) P_{ee}^{(2,2)}(z)
\end{equation}
This quantity is illustrated in figure~2 numerically. 
For $Q^2 \gsim 10 \GeV^2$, $P(z,Q^2;e^- \rightarrow e^+)$
may reach values
of ${\cal O}(1)$ for $z \sim 10^{-4}$.
Depending
on the detector resolution the charge of the scattered lepton
in  deep inelastic $ep$ collisions
might only be measurable in part of the kinematical range.
If the charge of the recorded final state electron is ignored
in the measurement, also the contribution
\begin{equation}
\label{conv}
\frac{d^2 \sigma^{(2,e^- \rightarrow e^+)}}{dx dy} =
\int_{z_0}^1 dz
P(z,Q^2; e^- \rightarrow e^+)
 {\cal J}(x,y,z)
\frac{d^2 \sigma^{(0)}}{dx dy} \left|_{x=\hat{x},
y=\hat{y},s=\hat{s}} \right.
\end{equation}
has to be included in the radiative corrections.
The corresponding contributions are given for the different
cases of measurement separately below.

\section{Numerical results}
In the numerical calculation the quark 
distributions are parametrized in the DIS scheme using the 
MRS $D^-$ parton distributions
\cite{MRSD} to illustrate the correction functions 
$\delta_{i}(x,y)$ for the different type of
measurements.
Equivalent results are obtained using the recent 
MRSH~\cite{MRSH}, CTEQ~\cite{CTEQ2}, or  GRV~\cite{GRV}
distributions.

The value of $\hat{Q}^2$
for   the structure functions in the radiative orders
in (\ref{eqMA}) may become smaller than the usual lower bound
$Q_0^2 \sim 4 \GeV^2$ for which the chosen
parton parametrization
applies.
Because of Lorentz invariance the structure 
functions $F_{2,3}^i(x,Q^2)$
vanish with $Q^2 \rightarrow 0$. Various parametrizations of
this behaviour have been  proposed recently~\cite{LOWQ2}. Here,
for an
illustration, we will
adopt a multiplicative factor~\cite{LQ2EXP}
\begin{equation}
\times \left [ 1 - \exp(-A^2 \hat{Q}^2) \right ]~~~{\rm with}~~~
A^2 = 3.37 \GeV^{-2}.
\end{equation}
for all structure functions in the numerical calculations.

Figure~3 illustrates the ${\cal O}(\alpha^2 L^2)$ corrections for
the case of jet measurement for neutral current 
deep inelastic  $e^- p$
scattering. The corrections are positive and grow with $x$ and $y$.
Compared with the 1st order results~\cite{RCJ1} they diminish the
negative ${\cal O}(\alpha)$ results in size.
 In the low $x$ range $x \lsim 0.01$
the 2nd order corrections are found  to be less than 1 \%.
Still
smaller values are found for the 2nd order contribution due to
$e^- \rightarrow e^+$ conversion, which amount to relative corrections
$< 4 \cdot 10^{-4}$ only.

The 2nd order
corrections for jet measurement in the case of charged current
deep inelastic $e^- p$ scattering are the  same as those of
the corresponding neutral current reactions  for large $x$, because
this range is dominated by the soft and virtual contributions,
cf. figure~4.
For smaller values of $x \lsim 0.01$ they are smaller than 1 \% and
may become negative, as also the 1st order corrections change
sign for small $x$~\cite{RCJ1}.
The contributions due to $e^- \rightarrow e^+$ conversion are somewhat
larger than in the case of neutral current reactions.

For the measurement based on mixed variables, i.e. $Q^2$ measured
at the
leptonic and $y_{JB}$ at the hadronic vertex (see figure~5),
 the 2nd order corrections
to the neutral current cross section
turn out to be less than 1 \% in the range $y \gsim 0.2$. They become
smaller with descending $x$  values. For $x \sim 0.5$ they reach
10  \%  in the range of small $y$. Note that
contrary to the
jet measurement the contributions
due to $e^- \rightarrow e^+$ conversion become of comparable order
as the 2nd order $e^- \rightarrow e^-$ corrections for small $y$ in
the case of mixed variables.

The ${\cal O}(\alpha)$ and ${\cal O}(\alpha^2)$ corrections for the
neutral current deep inelastic scattering cross section for a 
measurement using the double angle method are shown in figure~6.
Both the 1st and 2nd order corrections are nearly flat in $y$.
The corrections become smaller with descending values of $x$ and are
smaller than 3 \% for $x \lsim 0.01$. Due to
 these properties the
double angle method appears to be an ideal choice to measure the
neutral current cross section
from the point of view of radiative corrections.
The contributions from $e^- \rightarrow e^+$ conversion are less
than $2 \cdot 10^{-4}$.

Figure~7 shows the
${\cal O}(\alpha)$ and ${\cal O}(\alpha^2)$ corrections for the
neutral current deep inelastic scattering cross section, where
the scattering angle $\theta_e$ of the electron ond $y_{JB}$ are
used to define the kinematical variables. In this case the corrections
grow rapidly in the range $y \approx {\cal A}/(2E_e)$ and become
larger than 100 \% for $y \gsim {\cal A}/(2E_e)$ already
for $x = 0.01$.
 Even larger corrections are obtained for higher $x$ values.
 This  is
caused by the shift of the rescaled variables 
$\hat{x} \rightarrow 0$ {\sf  and} $\hat{Q}^2 \rightarrow 0$ in the
vicinity of $z_0 \approx y$. The relative correction becomes smaller
for very low values of $x$, since the differential cross sections
at the rescaled variables and the non-rescaled variables become
less different in value there. The 2nd order corrections are negative.
The contributions due to $e^- \rightarrow e^+$ conversion  grow with
rising values of $x$ and amount to ${\cal O}(2 \%)$ of the Born cross
section for $x \sim 0.5$  and larger values of $y$. Because
in part of the kinematical range
the correction
$\delta_{NC}(x,y)$ is significantly influenced by
the behaviour of $d \sigma^{(0)}/ dx dy$ in the non--perturbative
range, a reliable calculation of
$\delta_{NC}(x,y)$ is only possible
knowing the behaviour $d \sigma^{(0)}/ dx dy$ in the range of $Q^2, x
\rightarrow 0$. Due to this the plot of
$\delta_{NC}(x,y)$ in figure~7 serves only  to show the
{\it principal} behaviour under some assumptions made for
$d \sigma^{(0)}/ dx dy$. If, on the other hand, the cross section
$d \sigma / dx dy$, which includes the radiative corrections, is
well understood with
respect to the experimental systematics,
one may later try to use the measurement based on $\theta_e$ and $y_{JB}$
as a method to unfold
$d \sigma^{(0)}/ dx dy$ in the non--perturbative range.

In summary, we found that the ${\cal O}(\alpha^2 L^2)$ corrections,
supplemented by the soft exponentiation,
for the neutral and charged current deep inelastic $ep$
scattering cross sections
amount to $\lsim 10 \%$ for the cases of jet measurement,
mixed variables, and the double angle method. These contributions are
of the order of the difference between
the complete ${\cal O}(\alpha)$ and  the
leading log result ${\cal O}(\alpha L)$. In the case of the measurement
based on $\theta_e$ and $y_{JB}$, 
contributions from the non--perturbative
range $x, Q^2 \rightarrow 0$ influence both the 1st and 2nd order
corrections.
\newline
{\bf Acknowledgement.} I would like to thank Paul S\"oding for reading
the manuscript.
\newpage
%

%

\vspace{2cm}
\noindent
\begin{table}[hb]\centering
\begin{tabular}{|l|c|c|c|c|c|}
\hline
\multicolumn{1}{|l|}{            } &
\multicolumn{1}{ c|}{ $\hat{s}$  } &
\multicolumn{1}{ c|}{ $\hat{Q}^2$} &
\multicolumn{1}{ c|}{ $\hat{y}$  } &
\multicolumn{1}{ c|}{ $z_0$  } &
\multicolumn{1}{ c|}{ ${\cal J}(x,y,z)$} \\
\hline \hline
jet measurement     & $z s$ & $Q^2(1 - y)/(1 - y/z)$ & $y/z$ &
$y/[1 - x(1 - y)]$ &
$(1 - y)/(z -y)$    \\
mixed variables     & $z s$ & $Q^2 z$     & $y/z$ & $y$& 1     \\
double angle method & $z s$ & $Q^2 z^2$  & $y$ & 0 & $z$ \\
$y_{JB}$ and $\theta_e$ & $z s$ & $Q^2 z (z - y) / (1 - y)
$  & $y/z$ & $y$ &$ (z - y)/(1 - y)$ \\
\hline
\end{tabular}
\caption[xxx]{Description of parameters in eq.~(\ref{eqMA})
 for different types of cross section
measurement}
\label{tab-1}
\end{table}
\newpage

\vspace*{5cm}

\begin{picture}(120,20)(20,100) \centering
\put(0,-40){\epsfig{file=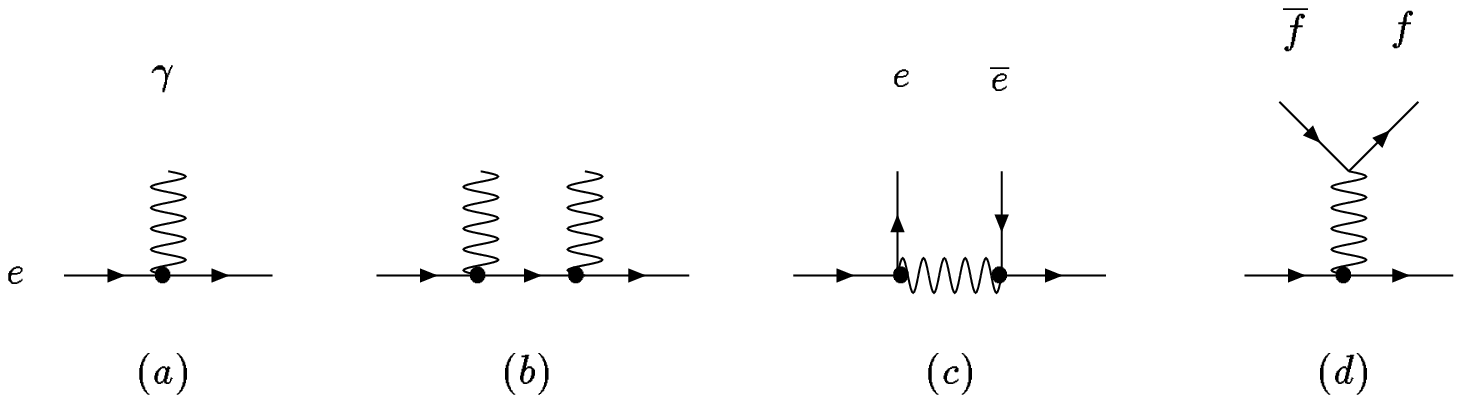,width=\textwidth}}
\end{picture}

\vspace{2mm}
\noindent
\small
\begin{center}
Figure~1:~Diagrams contributing to the radiative corrections
up to ${\cal O}(\alpha^2 L^2)$.
\end{center}
\normalsize
\begin{center}

\mbox{\epsfig{file=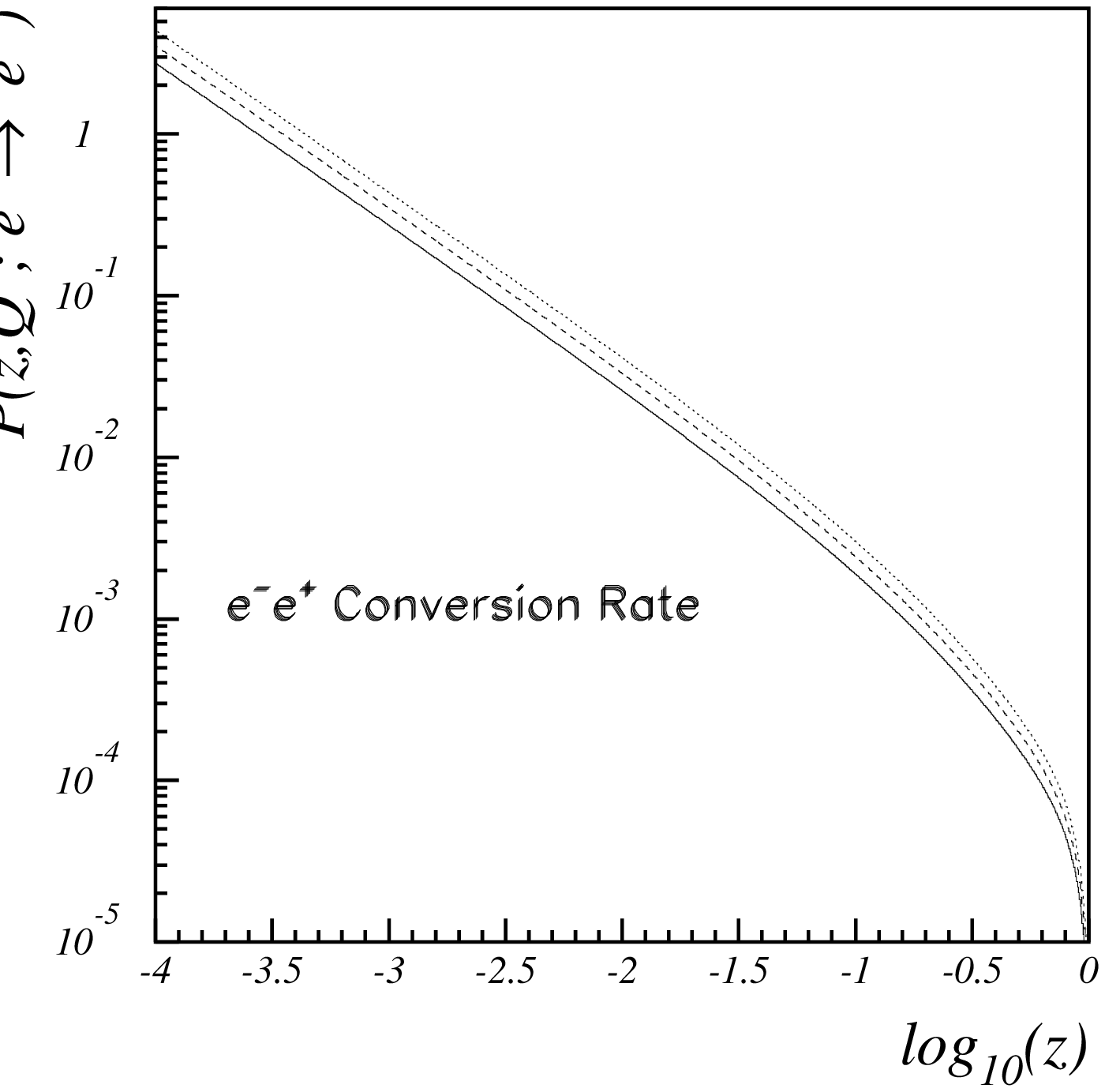,height=12cm,width=12cm}}

\vspace{2mm}
\noindent
\end{center}
\small
Figure~2:~$e^- \rightarrow e^+$ transition rate for
different values of $Q^2$. Full line: $Q^2 = 10 \GeV^2$,
dashed line: $Q^2 = 100 \GeV^2$, and
dotted line: $Q^2 = 1000 \GeV^2$.
\normalsize
\newpage
\begin{center}

\vspace{-.4in}

\mbox{\epsfig{file=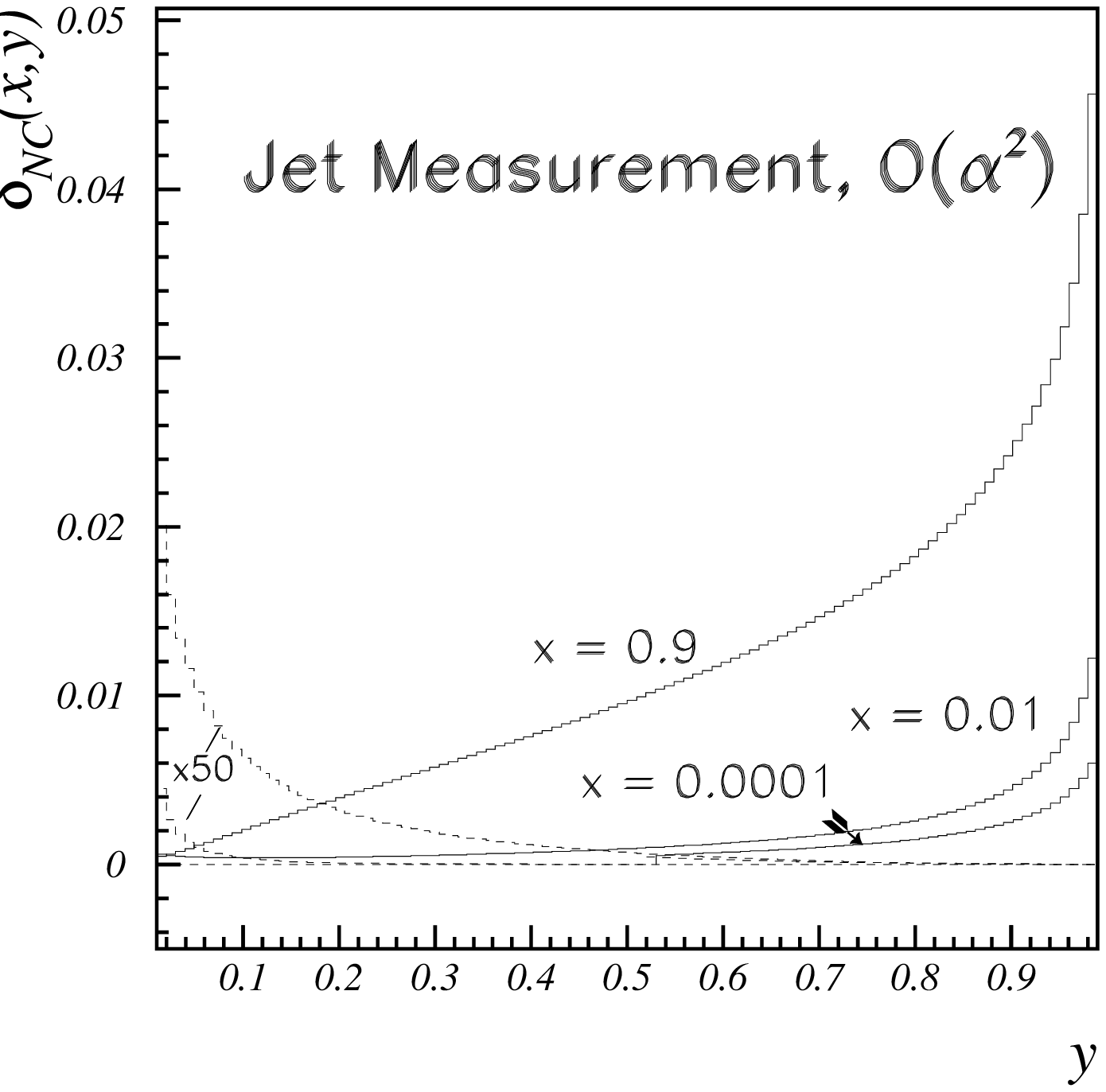,height=18cm,width=18cm}}
\end{center}

\vspace{2mm}
\noindent
\small
Figure~3:~Leptonic initial state  radiative corrections
$\delta_{NC}(x,y)$ = $(d \sigma^{(2 + >2, soft)}/dx dy)
/(d \sigma^{0}/dx dy)$  in LLA
for $e^- p$ deep inelastic scattering
in the case of jet measurement for $\sqrt{s} = 314 \GeV$,
${\cal A} = 0$, and $Q^2 \geq 5 \GeV^2$.
Full lines: 
${\cal O}(\alpha^2)$ corrections. Dashed lines: contributions
due to $e^- \rightarrow e^+$ conversion eq. (\ref{conv}),
$\delta_{NC}^{e^- \rightarrow e^+}(x,y)$ 
= $(d \sigma^{(2,e^- \rightarrow e^+)}/dx dy)
/(d \sigma^{0}/dx dy)$ scaled by $\times 50$; upper line:
$x = 0.01$, middle line: $x = 0.0001$, lower line $x = 0.9$.
\normalsize
\newpage
\begin{center}
\vspace{-.4in}

\mbox{\epsfig{file=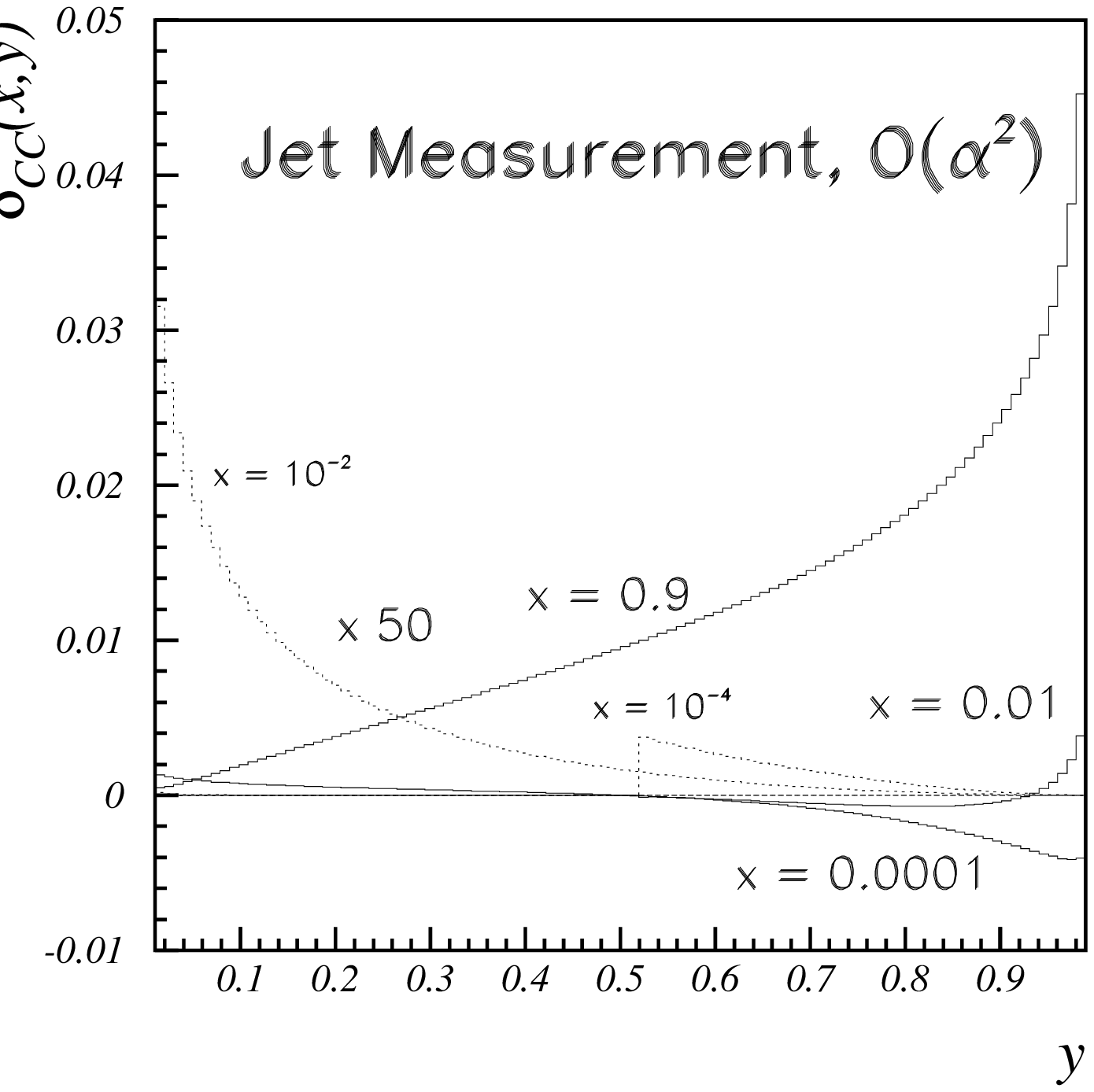,height=18cm,width=18cm}}
\end{center}

\vspace{2mm}
\noindent
\small
Figure~4:~$\delta_{CC}(x,y)$ = $(d \sigma^{(2 + >2, soft)}_{CC}/dx dy)
/(d \sigma^{0}_{CC}/dx dy)$ for deep inelastic $e^- p$
scattering
in the case of jet measurement. Dotted lines:
$\delta_{CC}^{e^- \rightarrow e^+}(x,y)$.
The  other
parameters are
the same as in figure~3.
\newpage
\begin{center}
\vspace{-.4in}

\mbox{\epsfig{file=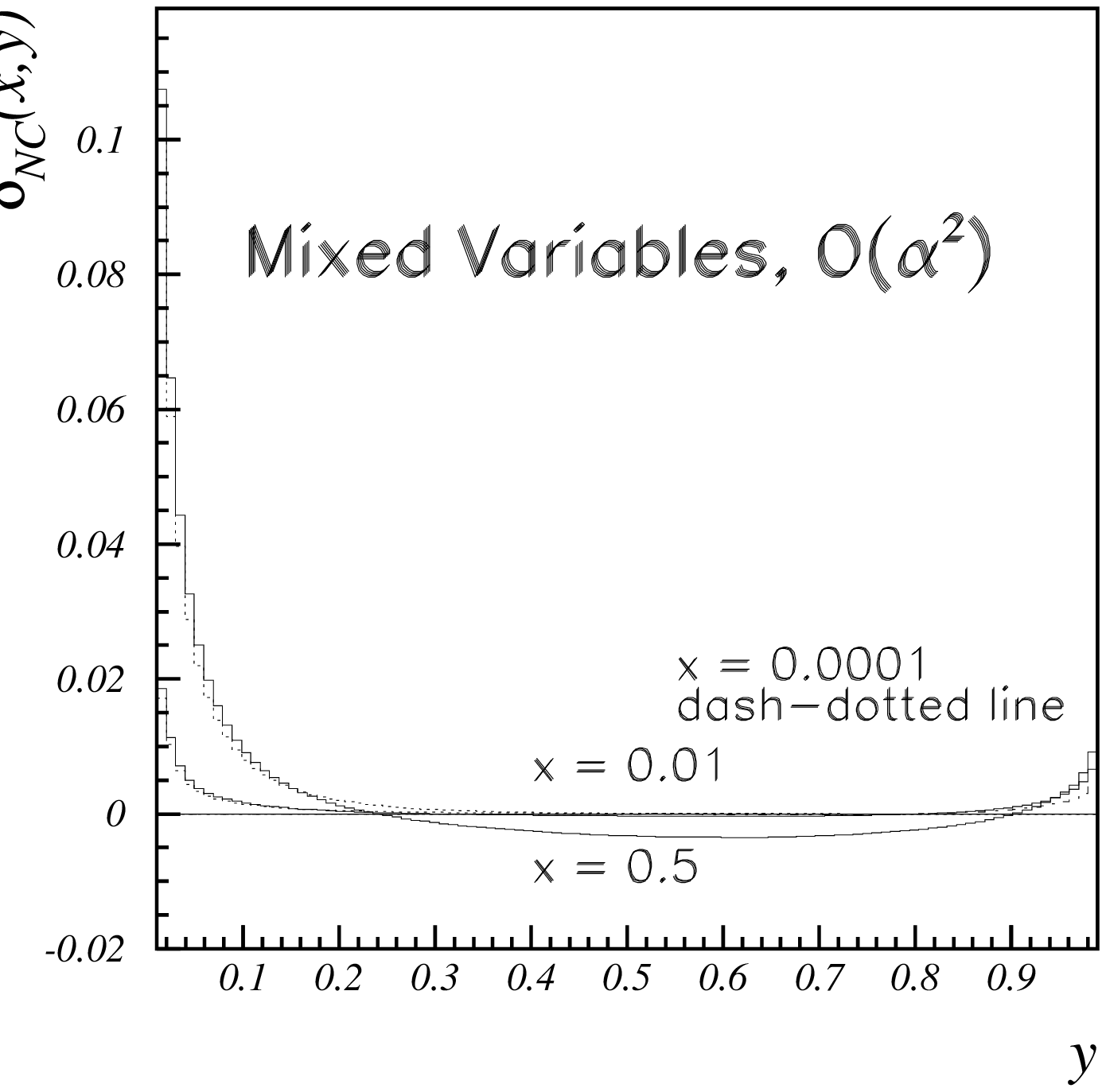,height=18cm,width=18cm}}
\end{center}

\vspace{2mm}
\noindent
\small
Figure~5:~$\delta_{NC}(x,y)$
for the case of mixed variables.
Dotted lines: $\delta_{NC}^{e^- \rightarrow e^+}(x,y)$; upper line:
$x = 0.5$, lower line $x = 0.01$.
The other  parameters are the same as in figure~3.

\normalsize
\newpage
\begin{center}
\vspace{-.4in}

\mbox{\epsfig{file=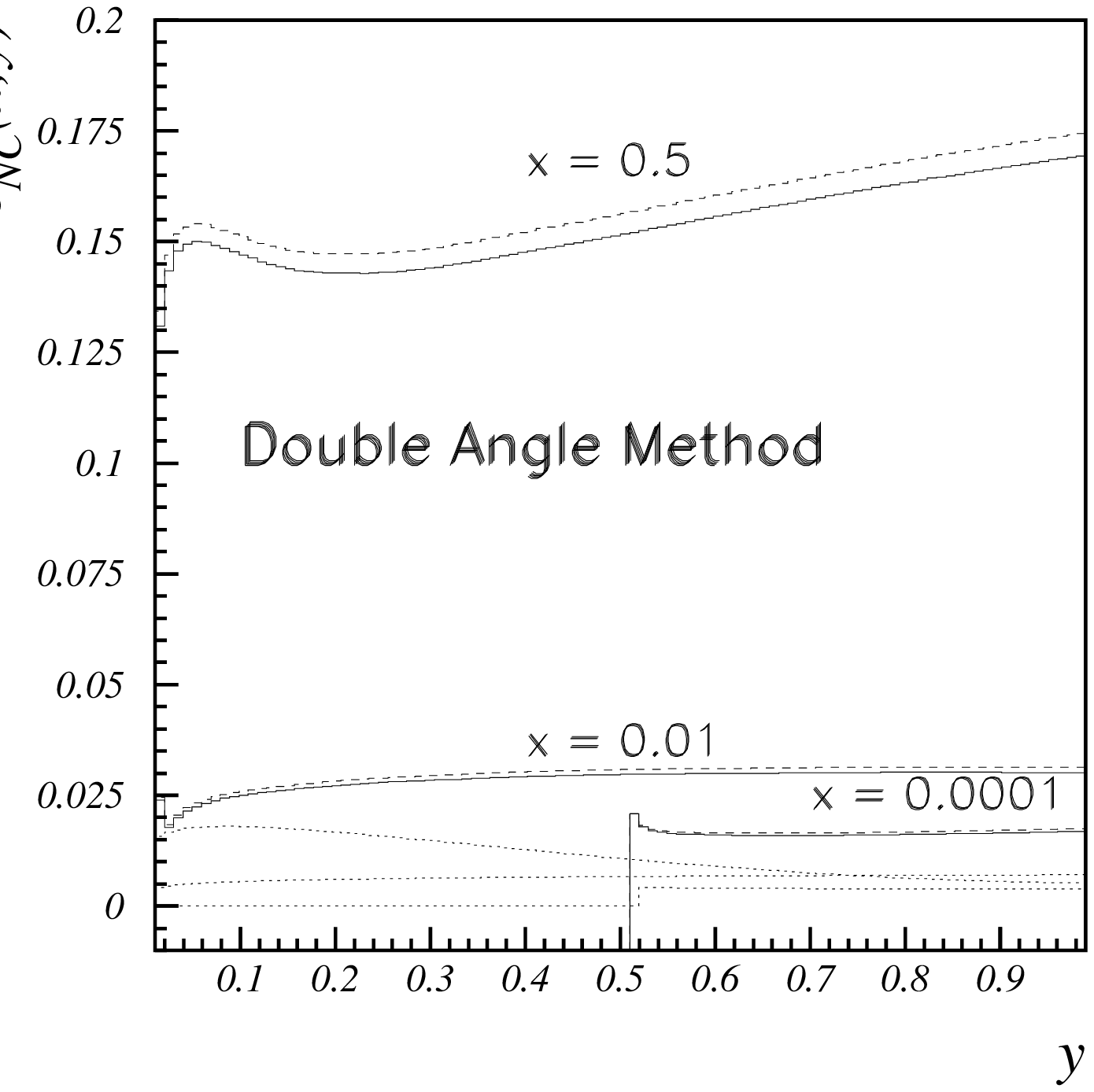,height=18cm,width=18cm}}

\vspace{2mm}
\noindent
\small
\end{center}
Figure~6:~$\delta_{NC}(x,y)$
for the case of the double angle method for
${\cal A} = 35 \GeV$.
Full lines: $\delta_{NC}^{(1+2 + >2, soft)}(x,y)$, dashed lines:
$\delta_{NC}^{(1)}(x,y)$.
Dotted lines:
$\delta_{NC}^{e^- \rightarrow e^+}(x,y)$ scaled by $\times 100$;
upper line: $x = 0.5$, middle line: $x = 0.01$, lower line:
$x = 0.0001$.
The other parameters are the same as in figure~3.
\newpage
\begin{center}
\vspace{-.4in}

\mbox{\epsfig{file=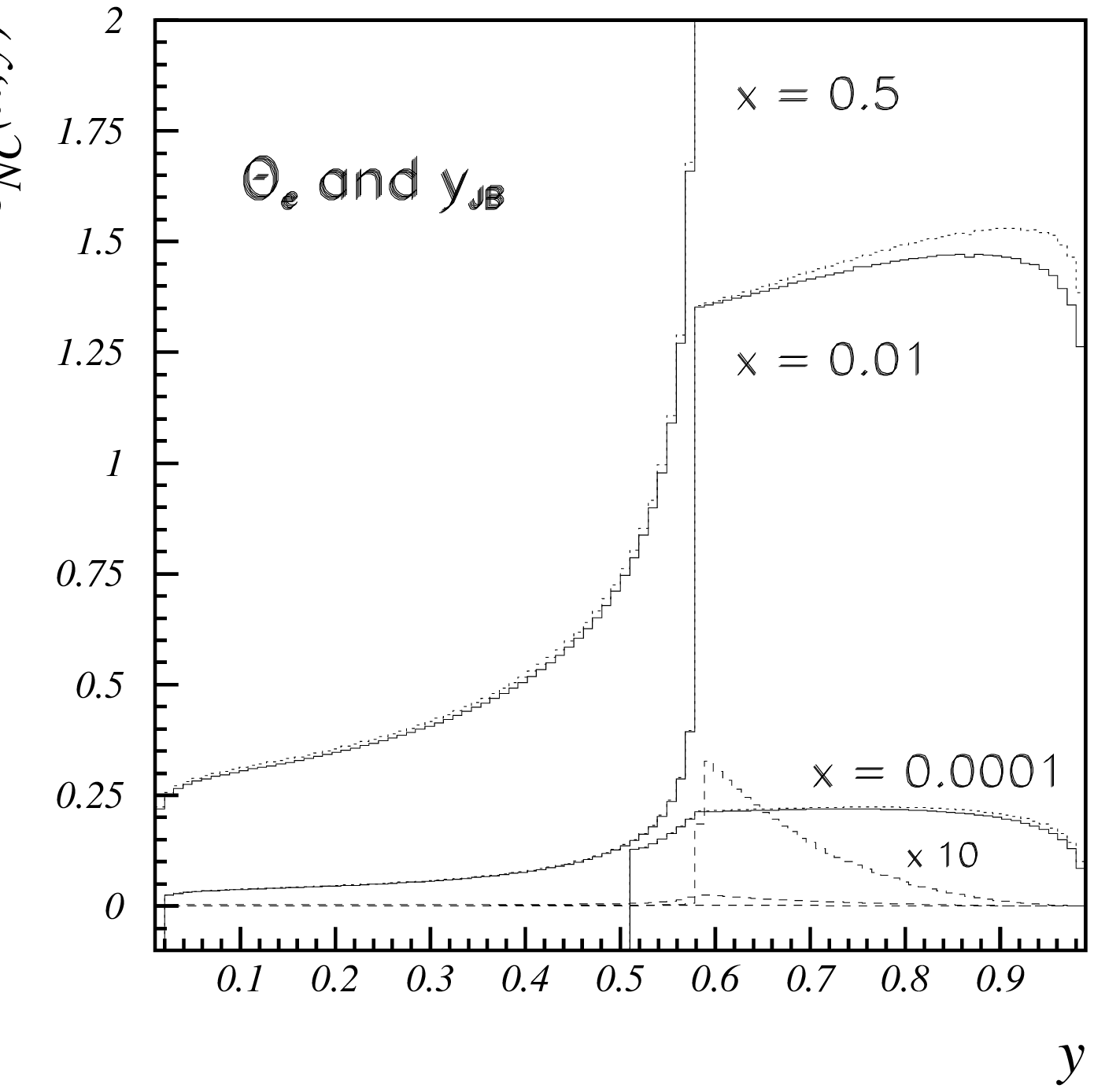,height=18cm,width=18cm}}

\vspace{2mm}
\noindent
\small
\end{center}
Figure~7:~$\delta_{NC}(x,y)$
for the measurement based on $\theta_e$ and $y_{JB}$ for
${\cal A} = 35 \GeV$.
Full lines: $\delta_{NC}^{(1+2 + >2, soft)}(x,y)$, dotted lines:
 $\delta_{NC}^{(1)}(x,y)$.
Dashed lines:
$\delta_{NC}^{e^- \rightarrow e^+}(x,y)$; upper line: $x = 0.5$,
middle line: $x = 10^{-2}$, lower line: $x = 10^{-4}$.
The other parameters are the same as in figure~3.
\normalsize
\end{document}